\documentclass[prd,showpacs,twocolumn,preprintnumbers,amsmath,amssymb,superscriptaddress,floatfix,nofootinbib]{revtex4-2}%

\usepackage{graphicx}
\usepackage{bm}
\usepackage{amsmath}
\usepackage{amsfonts}
\usepackage{amssymb}
\usepackage{color}
\usepackage{multirow}
\usepackage{subfigure}
\usepackage[colorlinks, citecolor=blue,anchorcolor=red,menucolor=red, linkcolor=red,filecolor=red,runcolor=red,urlcolor=blue,frenchlinks=red]{hyperref}

\begin{document}
	\title{Role of $a_0(1710)$ in the $J/\psi\to\rho^+\rho^-\omega$ and $J/\psi\to\gamma\rho^0\omega$ reactions}

	\author{Wen-Tao Lyu}\email{lvwentao9712@163.com}
	\affiliation{Departamento de Física Teórica and IFIC, Centro Mixto Universidad de Valencia-CSIC Institutos de Investigación de Paterna, 46071 Valencia, Spain}
	\affiliation{School of Physics, Zhengzhou University, Zhengzhou 450001, China}
	\vspace{0.5cm}
		
    \author{Luis Roca}\email{luisroca@um.es}
    \affiliation{Departamento de F\'isica, Universidad de Murcia, E-30100 Murcia, Spain}\vspace{0.5cm}
	
	\author{Eulogio Oset}\email{oset@ific.uv.es}
	\affiliation{Departamento de Física Teórica and IFIC, Centro Mixto Universidad de Valencia-CSIC Institutos de Investigación de Paterna, 46071 Valencia, Spain}
	\affiliation{Department of Physics, Guangxi Normal University, Guilin 541004, China}
	\vspace{0.5cm}

\begin{abstract}
We investigate the strong decay $J/\psi\to\rho^+\rho^-\omega$ and the radiative decay $J/\psi\to\gamma\rho^0\omega$, taking into account the $S$-wave $K^*\bar{K}^*$, $\rho \omega$ and $\rho \phi$ final-state interactions, which dynamically generate the scalar meson $a_0(1710)$. Our results demonstrate that a clear peak structure emerges around 1.8~GeV in the $\rho^+\omega$~($\rho^-\omega$) invariant mass distribution of the strong decay, which can be associated with the $a_0(1710)$ resonance. Similarly, a distinct peak is predicted in the $\rho^0\omega$ invariant mass distribution of the radiative decay. Our results indicate that clear signals of $a_0(1710)$ production could be observed in future measurements of these processes at BESIII, Belle II, and the planned Super Tau-Charm Facility~(STCF), thereby helping to determine its mass and width more precisely.

\end{abstract}
	
	\date{\today}
	
	\maketitle
	
\section{Introduction}\label{sec1}

Although the conventional quark–antiquark picture successfully describes most observed mesons, it faces difficulties in accounting for several states with nonconventional properties. The scalar-meson sector is particularly rich in such candidates, including $a_0(980)$, $f_0(980)$, $f_0(500)$, $f_0(1370)$, $f_0(1710)$ and $a_0(1710)$. The identification of light scalar mesons remains particularly challenging because of their large decay widths, and their internal structures are still actively debated. Several interpretations have been proposed for these states, including 
conventional $q\bar q$ configurations, multiquark states, hadronic molecules, glueballs, 
and mixtures of different components~\cite{Close:2002zu,Amsler:2004ps,Bugg:2004xu,Klempt:2007cp,Pelaez:2015qba,Nieves:1998hp,Janssen:1994wn,Wolkanowski:2015lsa,Jaffe:1976ig,Jaffe:2007id}, and reviews on the topic can be found in~\cite{Esposito:2016noz,Guo:2017jvc,Olsen:2017bmm,Lebed:2016hpi,Chen:2016qju,Liu:2019zoy,Brambilla:2019esw,Ali:2017jda,Karliner:2017qhf,Guo:2013sya,Wu:2022ftm}.

The $a_0(1710)$ resonance, the isospin partner of the scalar meson $f_0(1710)$, has only recently received direct experimental support. In 2022, the {\it BABAR} Collaboration first reported evidence for a scalar resonance, denoted as $a_0(1710)$, in the $\pi^\pm\eta$ invariant-mass spectrum of the process $\eta_c\to \eta\pi^+\pi^-$~\cite{BaBar:2021fkz}. Soon afterwards, the BESIII Collaboration reported evidence for this state in the $K_S^0K_S^0$ invariant-mass spectrum of $D_s^+ \to K_S^0 K_S^0\pi^+$~\cite{BESIII:2021anf} and in the $K_S^0K^+$ invariant-mass spectrum of $D_s^+ \to K_S^0 K^+\pi^0$~\cite{BESIII:2022npc}. It is worth noting that in Ref.~\cite{BESIII:2021anf}, BESIII did not distinguish between $a_0(1710)$ and $f_0(1710)$ in the $D_s^+ \to K_S^0 K_S^0\pi^+$ channel, and referred to the observed structure collectively as $S(1710)$. In Ref.~\cite{BESIII:2022npc}, the structure was labeled $a_0(1817)$ because the fitted Breit-Wigner mass was significantly higher. More recently, in 2023, the LHCb Collaboration confirmed this resonance in a Dalitz-plot analysis of $\eta_c$ decays, where it was denoted as $a_0(1700)$, and its resonance parameters were extracted~\cite{LHCb:2023evz}. The experimental determinations of the mass and width of $a_0(1710)$ are summarized in Table~\ref{table:mass_and_width_experiment}.

\begin{table}[htbp]
	\begin{center}
		\caption{ Experimental measurements on the mass ($M_{a_0(1710)}$) and width ($\Gamma_{a_0(1710)}$) of the scalar state $a_0(1710)$. The first error is statistical, and the second one is systematic. All values are in units of MeV.}		
		\begin{tabular}{lccc}
			\hline\hline
			Collaboration~~&~~Ref.~~&~~~~$M_{a_0(1710)}$~~~~&~~~~$\Gamma_{a_0(1710)}$~~~~\\
			\hline
			$\textit{BABAR}$ & \cite{BaBar:2021fkz} & $1704 \pm 5 \pm 2$ & $110 \pm 15 \pm 11$ \\
			BESIII & \cite{BESIII:2021anf} & $1723 \pm 11 \pm 2$ & $140 \pm 14 \pm 4$ \\
			BESIII & \cite{BESIII:2022npc} & $1817 \pm 8 \pm 20$ & $97 \pm 22 \pm 15$ \\
			LHCb & \cite{LHCb:2023evz} & $1736 \pm 10 \pm 12$ & $134 \pm 17 \pm 61$ \\
			\hline\hline
		\end{tabular}
		\label{table:mass_and_width_experiment}
	\end{center}
\end{table}

From the theoretical side, the $a_0(1710)$ state was predicted within the framework of the chiral unitary approach, as a molecular state dynamically generated by vector-meson interactions~\cite{Geng:2008gx}, using the local hidden gauge approach~\cite{Bando:1984ej,Bando:1987br,Meissner:1987ge,Nagahiro:2008cv} as a source of the interaction. This state appeared when the formalism used in the study of the $\rho\rho$ interaction~\cite{Molina:2008jw} was generalized to SU(3), where the $a_0(1710)$ appeared together with the $f_0(1710)$, the latter coupling mostly to $\rho\rho$, whereas the former couples mainly to $K^*\bar{K}^*$. In this picture, one considers coupled channels made of $\rho\rho$, $\rho\omega$, $\rho\phi$ and $K^*\bar{K}^*$ with different quantum numbers and the diagonal and transition potentials $V_{ij}$ between these channels, which are mediated by the exchange of vector mesons, are evaluated. Then $V_{ij}$ is used as the kernel of the Bethe-Salpeter equation, and the scattering matrices, $T = \left[1-VG\right]^{-1}V$, with $G = \text{diag}\left[G_i\right]$, and $G_i$ the meson-meson loop function for each channel, are evaluated, and one looks for poles in the complex plane. Close to these poles the scattering matrix behaves as $T_{ij} \sim \frac{g_i g_j}{s - s_R}$, where $s_R$ is the position of the complex pole. The relative strength of the couplings $g_i$ to the different channels provide information on the weight of each channel in the wave function of the state.

Besides the molecular interpretation discussed below, other scenarios have also been proposed for the $a_0(1710)$. Within a conventional quark-model framework, it has recently been argued that the experimental observation of the $a_0(1710)$ supports its assignment as an excited isovector $q\bar q$ state~\cite{Afonin:2025yfx}. The existence of an isovector scalar state in this mass region has also attracted attention in connection with the long-standing discussion on the nature of the $f_0(1710)$ and its possible glueball content, since the identification of an $a_0(1710)$ may have important implications for scalar-meson spectroscopy and glueball interpretations~\cite{Janowski:2014ppa,Guo:2022xqu}. At present, no consensus exists regarding the dominant structure of the $a_0(1710)$, and further theoretical and experimental studies are needed to clarify its nature.

An $a_0$ state in the same energy region was also obtained using dispersion relations~\cite{Du:2018gyn}. In both~\cite{Geng:2008gx} and~\cite{Du:2018gyn}, the mass of the state is, however, obtained at a higher value, around 1780 MeV, in between the masses determined in different experiments shown in Table~\ref{table:mass_and_width_experiment}. Later, it has been shown that the vector-vector picture remains essentially unchanged after including pseudoscalar-pseudoscalar coupled channels~\cite{Wang:2022pin}. Actually, in Ref.~\cite{Garcia-Recio:2010enl}, using an SU(6) scheme that mixes pseudoscalar-pseudoscalar and vector-vector channels, an $a_0$ state with pole position $\sqrt{s_R}=(1760-12i)$~MeV, (corresponding to a width of 24 MeV), was also predicted, coupling strongly to $K^*\bar{K}^*$ and $\phi\rho$. The vector-vector picture was also used in Ref.~\cite{Abreu:2023xvw}, but it is not the only possible one to produce an $a_0$ state around this energy, and in Ref.~\cite{Wang:2017pxm} mixing elements of quark model and Regge phenomenology, a state is obtained with mass around 1744 MeV. A similar classification using Regge trajectories is done in Ref.~\cite{Guo:2022xqu}.
\begin{table}[htbp]
	\begin{center}
		\caption{Couplings for the different channels. All values are in units of MeV.}
		\begin{tabular}{c|cccc}
			\hline\hline
			Pole & \multicolumn{4}{c}{$1780-i66$} \\
			\hline
			Channel & $K^*\bar{K}^*$ & $\rho\rho$ & $\rho\omega$ & $\rho\phi$ \\
			\hline
			$g$ & $7525-i1529$ & $0$ & $-4042+i1391$ & $4998-i1872$ \\
			\hline\hline
		\end{tabular}\label{table:coupling_geng}
	\end{center}
\end{table}

As summarized in Table~\ref{table:mass_and_width_experiment}, the experimental situation regarding the mass of the $a_0(1710)$ is still not settled, with different measurements not fully compatible within uncertainties, which in turn complicates the interpretation of its underlying structure. A particularly noteworthy aspect of Table~\ref{table:mass_and_width_experiment} is the BESIII result of Ref.~\cite{BESIII:2022npc}, which yields a mass around 1817 MeV, significantly larger than the values reported by BABAR, LHCb, and the BESIII analysis of Ref.~\cite{BESIII:2021anf}. This discrepancy has motivated some authors to refer to the structure observed in Ref.~\cite{BESIII:2022npc} as the $a_0(1817)$ rather than the $a_0(1710)$~\cite{Guo:2022xqu}, while other works denote it as the $a_0(1710)$ [$a_0(1817)$]~\cite{Oset:2023hyt}. However, it is worth noting that the relativistic Breit-Wigner parametrization employed in Ref.~\cite{BESIII:2022npc} may not be optimal for a broad resonance coupled to several channels, such as the $a_0(1710)$, and could lead to a biased determination of its mass. Improved analyses are currently in progress, and further experimental information will be needed to clarify this issue~\cite{baicianke_new}.

It has been shown in Refs.\cite{Zhu:2022wzk,Zhu:2022guw,Dai:2021owu,Wang:2023aza} that if the $a_0(1710)$ is treated as a $K^*\bar{K}^*$ molecular state, one is able to reproduce the invariant-mass distributions measured by the BESIII Collaboration in the processes $D_s^+ \to K_S^0 K_S^0\pi^+$ and $D_s^+ \to K_S^0 K^+\pi^0$~\cite{BESIII:2021anf,BESIII:2022npc}. Moreover, Ref.~\cite{Ding:2023eps} identified a pronounced dip structure around 1.8 GeV in the $\bar{K}^0K^+$ invariant-mass distribution in the $\eta_c \to \bar{K}^0 K^+ \pi^-$, associated with $a_0(1710)$, and achieved a successful description of the {\it BABAR} data. A perspective on the role played by the $a_0(1710)$ in these reactions, along with suggestions for observing it in new channels, is provided in Ref.~\cite{Oset:2023hyt}.

Hadronic decays of charmonium provide a valuable window into hadron-hadron interactions and therefore offer an important testing ground for quantum chromodynamics (QCD)~\cite{Dai:2026zqn,Lyu:2026rsm,Ding:2024lqk,Ding:2023eps,Liang:2019jtr}. In the present work, we study the strong and radiative decays of the $J/\psi$ by taking into account the $S$-wave vector-vector final state interaction, from which the scalar state $a_0(1710)$ is dynamically generated. The couplings of this resonance to the relevant channels are listed in Table~\ref{table:coupling_geng}~\cite{Geng:2008gx}. As can be seen, the $a_0(1710)$ couples strongly to $K^*\bar{K}^*$, $\rho\omega$ and $\rho\phi$, although its mass lies below the thresholds of the $K^*\bar{K}^*$ and $\rho\phi$ channels. Motivated by this feature, we propose to search for this resonance in the reactions $J/\psi\to\rho^+\rho^-\omega$ and $J/\psi\to\gamma\rho^0\omega$. Including the contribution from the $a_0(1710)$, we evaluate the $\rho^+\omega$ and $\rho^0\omega$ invariant-mass distributions in the strong and radiative decay modes, respectively. A further motivation to do this work is its likely observation in future BESIII experiments~\cite{baicianke}.

The remainder of this paper is organized as follows. In Sec.~\ref{sec2}, we present the theoretical formalism for the reactions $J/\psi\to\rho^+\rho^-\omega$ and $J/\psi\to\gamma\rho^0\omega$. The numerical results and corresponding discussions are given in Sec.~\ref{sec3}. Finally, a brief summary is provided in Sec.~\ref{sec4}.

\section{Formalism}\label{sec2}

Our theoretical framework begins with the premise that the $J/\psi$ meson is an SU(3) flavor singlet, as it is a $c\bar{c}$ state containing no $u$, $d$, or $s$ quarks. To describe its decay into three vector mesons, we construct invariant amplitudes by contracting scalars with three vector fields. This is achieved by evaluating the traces with the vector meson matrix $V$ of Eq.~(\ref{eq:Vmatrix}). There are three independent structures: $\langle VVV \rangle$, $\langle VV \rangle\langle V \rangle$ and $\langle V \rangle\langle V \rangle\langle V \rangle$. According to the principles of heavy quark spin symmetry~\cite{Manohar:1998xv,Abreu:2023yvf}, the dominant contributions are expected to arise from structures with fewer traces. This formalism is analogous to that employed in previous works dealing with related reactions involving vector–vector interactions in heavy quarkonium decays~\cite{Liang:2016hmr,Debastiani:2016ayp,Jiang:2019ijx,Sakai:2019uig,Molina:2019wjj}. The vector meson matrix is defined as

\begin{equation}\label{eq:Vmatrix}
	V =
	\left(
	\begin{array}{ccc}
		\frac{1}{\sqrt{2}}\rho^0 + \frac{1}{\sqrt{2}}\omega  & \rho^+ & K^{*+} \\
		\rho^- & -\frac{1}{\sqrt{2}}\rho^0 + \frac{1}{\sqrt{2}}\omega  & ~K^{*0}~ \\
		K^{*-} & \bar{K}^{*0} & \phi  \\
	\end{array}
	\right).
\end{equation}

Expanding the traces, we obtain

\begin{align}
	\langle VVV\rangle
	={}&
	\phi^3 +\frac{1}{\sqrt{2}} \omega^3 + 3 K^{*+} K^{*-} \phi + \frac{3}{\sqrt{2}} K^{*+} K^{*-} \omega  \nonumber\\
	& + 3\sqrt{2} \omega \rho^+ \rho^- + 3 \rho^+ K^{*-} K^{*0} + \frac{3}{\sqrt{2}} K^{*+} K^{*-} \rho^0  \nonumber\\
	&  + \frac{3}{\sqrt{2}} \omega (\rho^0)^2 + 3 \rho^- K^{*+} \bar{K}^{*0}+ 3 K^{*0} \bar{K}^{*0} \phi \nonumber\\
	& + \frac{3}{\sqrt{2}} K^{*0} \bar{K}^{*0} \omega - \frac{3}{\sqrt{2}} K^{*0} \bar{K}^{*0} \rho^0 ,\label{eq:VVV}
\end{align}
\begin{align}
		\langle VV\rangle\langle V\rangle
		={}& 
		\phi^3 + \sqrt{2}\,\omega\phi^2 + \omega^2\phi + \sqrt{2}\,\omega^3  \nonumber\\
		& + 2K^{*+}K^{*-}\,\phi + 2\sqrt{2}\,\omega K^{*+}K^{*-} + 2\,\rho^+\rho^-\phi  \nonumber\\
		& + 2\sqrt{2}\,\omega\rho^+\rho^- + (\rho^0)^2\phi + \sqrt{2}\,\omega(\rho^0)^2  \nonumber\\
		& + 2\,K^{*0}\bar{K}^{*0}\phi + 2\sqrt{2}\,\omega K^{*0}\bar{K}^{*0} ,\label{eq:VV_V}
\end{align}
\begin{equation}
	\begin{aligned}
		\langle V\rangle\langle V\rangle\langle V\rangle
		=
		\phi^3+3\sqrt{2}\,\phi^2\,\omega+6\,\phi\,\omega^2+2\sqrt{2}\,\omega^3.
	\end{aligned}\label{eq:V_V_V}
\end{equation}
To evaluate the relative contributions of these three distinct structures, we introduce independent weight parameters $A$, $B$ and $C$. Thus, for practical purposes, we have
\begin{equation}\label{eq:three_structure_VVV}
	H=A\langle VVV\rangle+B\langle VV\rangle\langle V\rangle+C\langle V\rangle\langle V\rangle\langle V\rangle.
\end{equation}

\subsection{Strong decay of $J/\psi\to\rho^+\rho^-\omega$}
\begin{figure}[htbp]
	\centering
	
	\includegraphics[scale=0.55]{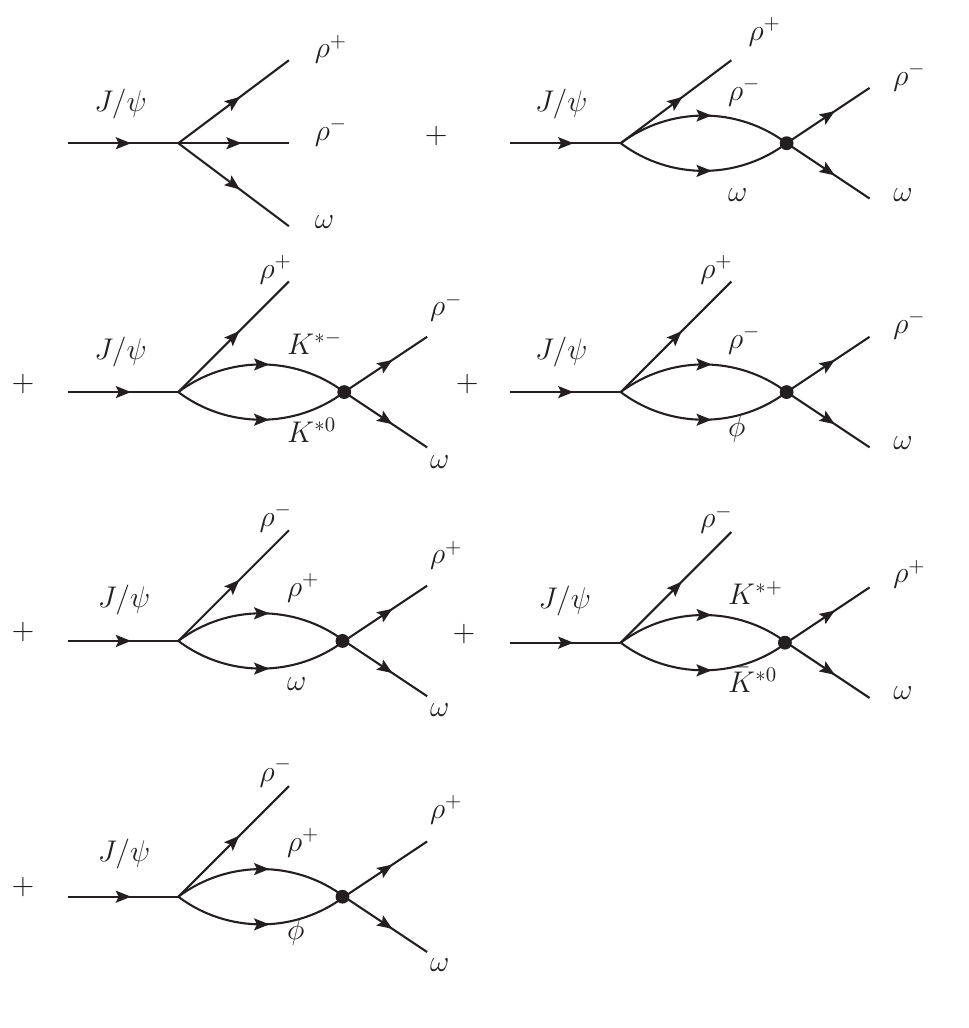}
	
	\caption{Mechanisms for tree level $J/\psi\to\rho^+\rho^-\omega$ and rescattering of intermediate components.}\label{fig:strong_decay_diagram}
\end{figure}

We first investigate the dynamics of the $J/\psi\to\rho^+\rho^-\omega$ decay, considering both tree-level mechanisms and final-state rescattering, the latter being responsible for the dynamical generation of the $a_0(1710)$. The complete set of mechanisms is illustrated in Fig.~\ref{fig:strong_decay_diagram}. To evaluate these diagrams, we first calculate the weights of the primary vertices, which determine their contribution to the amplitude, using Eqs.~(\ref{eq:VVV}), (\ref{eq:VV_V}) and (\ref{eq:V_V_V}), which yield
\begin{equation}
	\begin{aligned}
		\rho^+\rho^-\omega
		\Rightarrow{}&3\sqrt{2}\,A+2\sqrt{2}\,B, \\
		\rho^+K^{*-}K^{*0}\Rightarrow{}&3\,A, \\
		\rho^-K^{*+}\bar{K}^{*0}\Rightarrow{}&3\,A, \\
		\rho^+\rho^-\phi\Rightarrow{}&2\,B. 
	\end{aligned}\label{eq:strong_vertex}
\end{equation}
Note that there is no dependence on the $C$ coefficient of Eq.~(\ref{eq:three_structure_VVV}).

Furthermore, we must account for the spin dynamics of the vector-vector pairs. For a $VV \to VV$ transition, the amplitude involves four polarization vectors: $\vec{\epsilon}_1$, $\vec{\epsilon}_2$, $\vec{\epsilon}_3$ and $\vec{\epsilon}_4$ for $1+2\to3+4$. The corresponding spin projectors for total spin $S=0,1,2$ are given by~\cite{Molina:2008jw,Geng:2008gx}
\begin{equation}
	\begin{aligned}
		P^{(0)}={}&\frac{1}{3}
		(\vec{\epsilon}_1\cdot \vec{\epsilon}_2)
		(\vec{\epsilon}_3\cdot \vec{\epsilon}_4), \\
		P^{(1)}={}&\frac{1}{2}
		\left[
		(\vec{\epsilon}_1\cdot \vec{\epsilon}_3)
		(\vec{\epsilon}_2\cdot \vec{\epsilon}_4)
		-
		(\vec{\epsilon}_1\cdot \vec{\epsilon}_4)
		(\vec{\epsilon}_2\cdot \vec{\epsilon}_3)
		\right], \\
		P^{(2)}={}&\frac{1}{2}
		\left[
		(\vec{\epsilon}_1\cdot \vec{\epsilon}_3)
		(\vec{\epsilon}_2\cdot \vec{\epsilon}_4)
		+
		(\vec{\epsilon}_1\cdot \vec{\epsilon}_4)
		(\vec{\epsilon}_2\cdot \vec{\epsilon}_3)
		\right]\\
		&-\frac{1}{3}
		(\vec{\epsilon}_1\cdot \vec{\epsilon}_2)
		(\vec{\epsilon}_3\cdot \vec{\epsilon}_4).
	\end{aligned}
\end{equation}

\begin{figure}[htbp]
	\centering
	\includegraphics[scale=0.55]{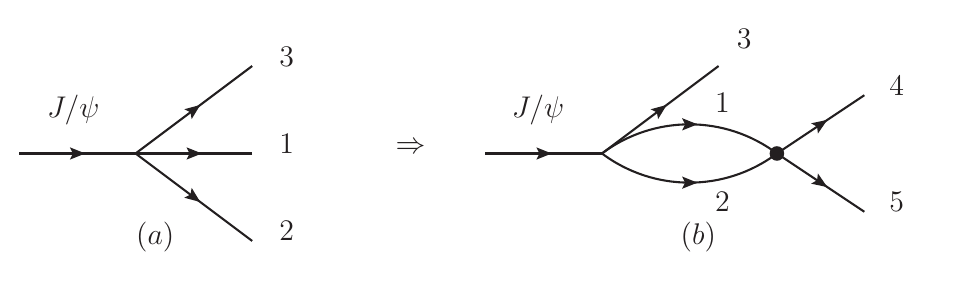}
	\caption{Topological structures of the polarization vectors. (a) tree level; (b) rescattering.}\label{fig:polarization_vectors_diagram_diagram}
\end{figure}

These processes involve two distinct topological structures, as depicted in Fig.~\ref{fig:polarization_vectors_diagram_diagram}. For the tree-level process (first diagram), the polarization structure is 
\begin{equation}\label{eq:polarization_structure_tree}
\epsilon_{J/\psi i}\,\epsilon_{3i}\,
\epsilon_{1j}\,\epsilon_{2j}
\;+\;\epsilon_{J/\psi i}\,\epsilon_{1i}\,
\epsilon_{2j}\,\epsilon_{3j}
\;+\;\epsilon_{J/\psi i}\,\epsilon_{2i}\,
\epsilon_{1j}\,\epsilon_{3j}.
\end{equation}
Considering that the $a_0(1710)$ is an $S=0$ state, we project the amplitude using $P^{(0)}$. Consequently, for the loop-level process (second diagram), the three structures of Eq.~(\ref{eq:polarization_structure_tree}) become
\begin{equation}
	\begin{aligned}
	\epsilon_{J/\psi i}\,\epsilon_{3i}\,\epsilon_{1j}\,\epsilon_{2j}\,t_{12,45}\,\frac{1}{3}\,\epsilon_{1l}\,\epsilon_{2l}\,\epsilon_{4m}\,\epsilon_{5m}\\
	=\epsilon_{J/\psi i}\,\epsilon_{3i}\,t_{12,45}\,\epsilon_{4m}\,\epsilon_{5m},\\
	\epsilon_{J/\psi i}\,\epsilon_{1i}\,\epsilon_{2j}\,\epsilon_{3j}\,t_{12,45}\,\frac{1}{3}\,\epsilon_{1l}\,\epsilon_{2l}\,\epsilon_{4m}\,\epsilon_{5m}\\
	=\epsilon_{J/\psi i}\,\epsilon_{3i}\,\frac{1}{3}\,t_{12,45}\,\epsilon_{4m}\,\epsilon_{5m},\\
	\epsilon_{J/\psi i}\,\epsilon_{2i}\,\epsilon_{1j}\,\epsilon_{3j}\,t_{12,45}\,\frac{1}{3}\,\epsilon_{1l}\,\epsilon_{2l}\,\epsilon_{4m}\,\epsilon_{5m}\\
	=\epsilon_{J/\psi i}\,\epsilon_{3i}\,\frac{1}{3}\,t_{12,45}\,\epsilon_{4m}\,\epsilon_{5m},
	\end{aligned}
\end{equation}
where we applied the polarization sum $\sum\epsilon_{i}\epsilon_{j}=\delta_{ij}$. The total transition matrix $t$, summing over all mechanisms in Fig.~\ref{fig:strong_decay_diagram}, can then be written as
\begin{equation}
	\begin{aligned}
		t=&
		\alpha \,\epsilon_{J/\psi i}\,\epsilon_{\rho^+i}\,\epsilon_{\rho^-j}\,\epsilon_{\omega j}\\
		&+\beta \,\epsilon_{J/\psi i}\,\epsilon_{\rho^-i}\,\epsilon_{\rho^+j}\,\epsilon_{\omega j}\\
		&+\gamma \,\epsilon_{J/\psi i}\,\epsilon_{\omega i}\,\epsilon_{\rho^+j}\,\epsilon_{\rho^-j},
	\end{aligned}
\end{equation}
with
	\begin{align}
		\alpha
		=&
		\left(3\sqrt{2}A+2\sqrt{2}B\right)
		\left(
		1+\frac{5}{3}\,G_{\rho^-\omega}\,t_{\rho^-\omega,\rho^-\omega}(M_{\text{inv}}(\rho^-\omega))
		\right)\nonumber\\
		&+3A\frac{5}{3}\,G_{K^{*-}K^{*0}}\,t_{K^{*-}K^{*0},\rho^-\omega}(M_{\text{inv}}(\rho^-\omega))\nonumber\\
		&+2B\frac{5}{3}\,G_{\rho^-\phi}\,t_{\rho^-\phi,\rho^-\omega}(M_{\text{inv}}(\rho^-\omega)),\label{eq:alpha}
	\end{align}
	\begin{align}
		\beta
		=&
		\left(3\sqrt{2}A+2\sqrt{2}B\right)
		\left(
		1+\frac{5}{3}\,G_{\rho^+\omega}\,t_{\rho^+\omega,\rho^+\omega}(M_{\text{inv}}(\rho^+\omega))
		\right)\nonumber\\
		&+3A\frac{5}{3}\,G_{K^{*+}\bar{K}^{*0}}\,t_{K^{*+}\bar{K}^{*0},\rho^+\omega}(M_{\text{inv}}(\rho^+\omega))\nonumber\\
		&+2B\frac{5}{3}\,G_{\rho^+\phi}\,t_{\rho^+\phi,\rho^+\omega}(M_{\text{inv}}(\rho^+\omega)),\label{eq:beta}
	\end{align}
\begin{equation}
	\begin{aligned}
		\gamma
		=
		3\sqrt{2}A+2\sqrt{2}B.
	\end{aligned}\label{eq:gamma}
\end{equation}
Here, $G$ denotes the standard meson-meson loop function, regularized using the cut-off method with $q_{\text{max}}=960$~MeV~\cite{Geng:2008gx}. The $G$ functions are logarithmically divergent and require regularization to render them finite. This is usually done using dimensional regularization, or cut-off regularization, where one takes a maximum three-momentum in the loop integration. The choice of this cut-off momentum, $q_{\text{max}}$, is not arbitrary, but is a parameter which is obtained from experiment. It is generally accepted that it is of the order of the vector meson masses, but the precise value is obtained by fitting the pole positions to the experimental masses as an average for the different observed states. This was done in~\cite{Geng:2008gx} yielding $q_{max} \simeq 960 \text{ MeV}$. This value is of natural size and was fixed in Ref.~\cite{Geng:2008gx} from a global description of the vector-vector interaction within the local hidden gauge approach, successfully reproducing several dynamically generated resonances, including the $a_0(1710)$. Since the present work employs the same vector-vector amplitudes and resonance parameters, we keep this value unchanged for consistency, but we shall make some variations at the end to estimate possible uncertainties from this source.

The transition amplitudes are given by
\begin{equation}
	t_{\rho^-\omega,\rho^-\omega}=\dfrac{g_{\rho\omega}g_{\rho\omega}}{M_{\text{inv}}^2(\rho^-\omega)-M_R^2+iM_R\Gamma_R},
\end{equation}
\begin{equation}
	t_{\rho^+\omega,\rho^+\omega}=\dfrac{g_{\rho\omega}g_{\rho\omega}}{M_{\text{inv}}^2(\rho^+\omega)-M_R^2+iM_R\Gamma_R},
\end{equation}
\begin{equation}
	t_{K^{*-}K^{*0},\rho^-\omega}=-\dfrac{g_{K^{*}\bar{K}^{*}}g_{\rho\omega}}{M_{\text{inv}}^2(\rho^-\omega)-M_R^2+iM_R\Gamma_R},
\end{equation}
\begin{equation}
	t_{K^{*+}\bar{K}^{*0},\rho^+\omega}=-\dfrac{g_{K^{*}\bar{K}^{*}}g_{\rho\omega}}{M_{\text{inv}}^2(\rho^+\omega)-M_R^2+iM_R\Gamma_R},
\end{equation}
\begin{equation}
	t_{\rho^-\phi,\rho^-\omega}=\dfrac{g_{\rho\phi}g_{\rho\omega}}{M_{\text{inv}}^2(\rho^-\omega)-M_R^2+iM_R\Gamma_R},
\end{equation}
\begin{equation}
	t_{\rho^+\phi,\rho^+\omega}=\dfrac{g_{\rho\phi}g_{\rho\omega}}{M_{\text{inv}}^2(\rho^+\omega)-M_R^2+iM_R\Gamma_R},
\end{equation}
where the couplings $g_i$ are listed in Table~\ref{table:coupling_geng}, and the resonance parameters are $M_R=1780$~MeV and $\Gamma_R=132$~MeV. The minus signs in the formulas arise from the isospin multiplets ($\bar{K}^{*0}$, $-K^{*-}$) and ($-\rho^+$, $\rho^0$, $\rho^-$). Summing and averaging $|t|^2$ over the particle spins yields
\begin{align}\label{eq:ttotal_strong}
	&\overline{\sum}\sum|t|^2
	=\frac{1}{3}\left\{9\left|\alpha\right|^2+9\left|\beta\right|^2+9\left|\gamma\right|^2\right.\nonumber\\
	&\left.+3\times2\textbf{Re}(\alpha\beta^*)+3\times2\textbf{Re}(\beta\gamma^*)+3\times2\textbf{Re}(\alpha\gamma^*)\right\}\nonumber\\
	&=\left\{3\left|\alpha\right|^2+3\left|\beta\right|^2+3\left|\gamma\right|^2+2\textbf{Re}(\alpha\beta^*)+2\textbf{Re}(\beta\gamma^*)\right.\nonumber\\
	&\left.+2\textbf{Re}(\alpha\gamma^*)\right\}.
\end{align}

Assigning the indices 1 to $\rho^-$, 2 to $\omega$ and 3 to $\rho^+$, we have
\begin{equation}
	\dfrac{d^2\Gamma}{dM_{12}dM_{23}} = \frac{1}{(2\pi)^3}\dfrac{1}{32m_{J/\psi}^3}\overline{\sum}\sum|t|^2~2M_{12}~2M_{23},
\end{equation}
using the Mandl and Shaw normalization for the meson fields~\cite{Mandl:1985bg}.

We can obtain $d\Gamma/dM_{12}$ by integrating $d^2\Gamma/(dM_{12}dM_{23})$ over $M_{23}$ within the kinematic limits. Permuting the indices allows us to evaluate all three invariant mass distributions. We use $M_{12}$ and $M_{23}$ as independent variables, and determine $M_{13}$ using the kinematic relation $M_{12}^2+M_{13}^2+M_{23}^2=m_{J/\psi}^2+m_{\rho^-}^2+m_{\omega}^2+m_{\rho^+}^2$ to get $M_{13}$ from them.

\subsection{Radiative decay of $J/\psi\to\gamma\rho^0\omega$}

\begin{figure}[htbp]
	\centering
	
	\includegraphics[scale=0.55]{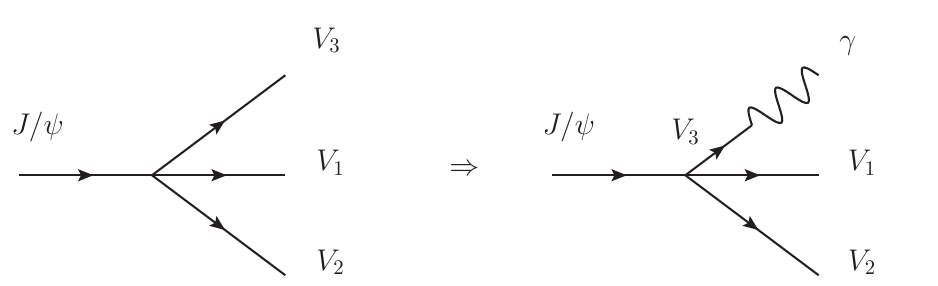}
	
	\caption{Diagrammatic representation of the Vector Meson Dominance (VMD) mechanism.}\label{fig:radiative_vector_diagram}
\end{figure}

Next, we turn our attention to the radiative decay $J/\psi\to\gamma\rho^0\omega$. To couple the primary $VVV$ vertex to a final photon, we employ the Vector Meson Dominance (VMD) model. In this approach, the $J/\psi$ initially decays into intermediate vector mesons ($\rho^0$, $\omega$, or $\phi$), which subsequently convert into a photon via the interaction Lagrangian~\cite{Bando:1984ej,Bando:1987br,Meissner:1987ge,Nagahiro:2008cv,Su:2025aiz}
\begin{equation}
	\begin{aligned}[b]\label{eq:L_Vgamma}
		\mathcal{L}_{V\gamma}=-\,M_{V}^2 \;\dfrac{e}{g}\; A_{\mu}\langle Q\, V^{\mu}  \rangle,
	\end{aligned}
\end{equation}
where $V$ is again the matrix of Eq.~(\ref{eq:Vmatrix}), $A_\mu$ the photon field and $e<0$ with $e^2=4\pi/137$ and $g=M_V/2f$, with $M_V$ the vector mass and $f$ the pion decay constant, $f=93$~MeV. In Eq.~\eqref{eq:L_Vgamma}, $Q$ is the quark charge matrix,
\begin{equation}
	Q = \frac{1}{3}
	\left( \begin{array}{ccc}
		2 & 0 & 0 \\
		0 & -1 & 0 \\
		0 & 0 & -1
	\end{array} \right).
\end{equation}
This conversion Lagrangian generates the vertex
\begin{equation}\label{eq:Lvertex}
	\begin{aligned}[b]
		-it&=-i\frac{M_{V}^{2}\,e}{g}\;A_{\mu}\; V^{\mu}\;C_{\gamma},
	\end{aligned}
\end{equation}
with the flavor factor $C_{\gamma}$ given by
\begin{equation}
	\begin{aligned}[b]
		C_{\gamma}=\begin{Bmatrix}
			\frac{1}{\sqrt{2}},~~\rho^0 
			\\
			\frac{1}{3}\frac{1}{\sqrt{2}},~\omega 
			\\
			-\frac{1}{3},~~\phi
			
		\end{Bmatrix}.
	\end{aligned}\label{eq:Cgamma}
\end{equation}

Thus, as illustrated in Fig.~\ref{fig:radiative_vector_diagram}, the replacement rule for photon emission is
\begin{equation}
	\begin{aligned}[b]
		&t^\prime=t_{V_3V_1V_2}\left(\epsilon_{J/\psi i}\,\epsilon_{3i}\,\epsilon_{1j}\,\epsilon_{2j}\;\right.  \\
		&\left.+\;\epsilon_{J/\psi i}\,\epsilon_{1i}\,\epsilon_{2j}\,\epsilon_{3j}\;+\;\epsilon_{J/\psi i}\,\epsilon_{2i}\,\epsilon_{1j}\,\epsilon_{3j}\right)    \\
		&~~~~~~~~~~~~~~~~~~~~~~\Downarrow                         \\
		&t^\prime=t_{V_3V_1V_2}\frac{e}{g}\;C_{\gamma}\left(\epsilon_{J/\psi i}\,\epsilon_{\gamma i}\,\epsilon_{1j}\,\epsilon_{2j}\;\right.  \\
		&\left.+\;\epsilon_{J/\psi i}\,\epsilon_{1i}\,\epsilon_{2j}\,\epsilon_{\gamma j}\;+\;\epsilon_{J/\psi i}\,\epsilon_{2i}\,\epsilon_{1j}\,\epsilon_{\gamma j}\right).
	\end{aligned}\label{eq:the_replacement_rule}
\end{equation}

\begin{figure}[htbp]
	\centering
	
	\includegraphics[scale=0.55]{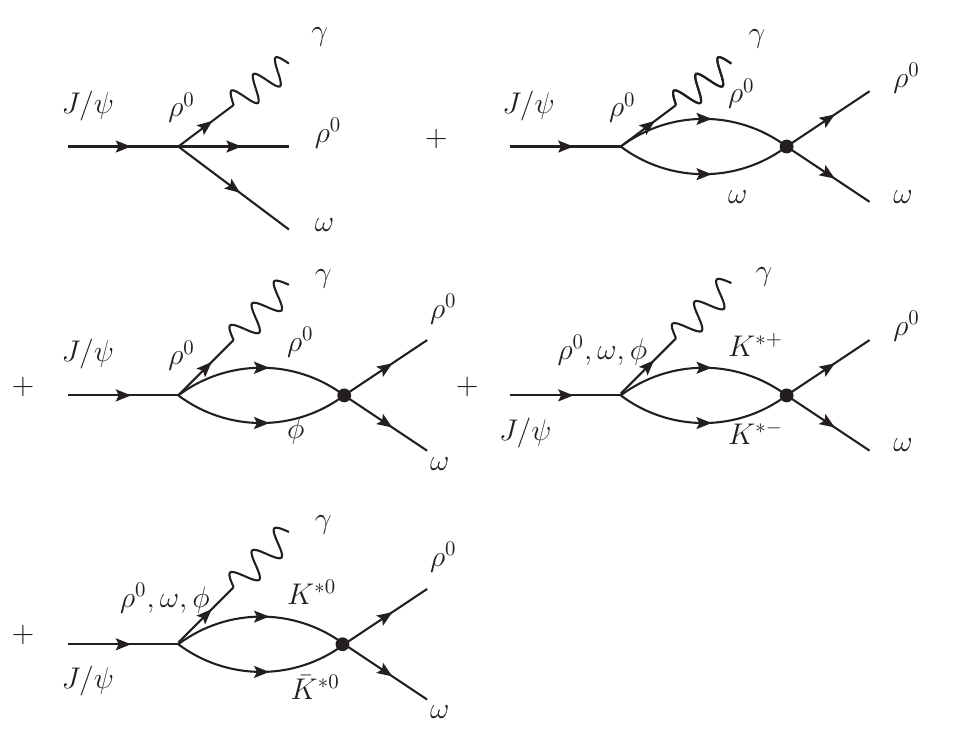}
	
	\caption{Mechanisms for tree level $J/\psi\to\gamma\rho^0\omega$ and rescattering of intermediate components.}\label{fig:radiative_decay_diagram}
\end{figure}

The complete set of mechanisms considered is presented in Fig.~\ref{fig:radiative_decay_diagram}. The primary vertex terms $t_{\gamma V_1V_2}C_{\gamma}$ can be evaluated using Eqs.~(\ref{eq:VVV}),~(\ref{eq:VV_V}),~(\ref{eq:V_V_V}) and~(\ref{eq:Cgamma}), yielding
\begin{equation}
	\begin{aligned}
		t_{\gamma\rho^0\omega}C_{\gamma}=&3A+2\,B, \\
		t_{\gamma\rho^0\phi}C_{\gamma}=&\sqrt{2}B, \\
		t_{\gamma K^{*+}K^{*-}}C_{\gamma}=&A, \\
		t_{\gamma K^{*0}\bar{K}^{*0}}C_{\gamma}=&-2A. 
	\end{aligned}\label{eq:radiative_vertex}
\end{equation}

Consequently, the total radiative amplitude reads
\begin{equation}
	\begin{aligned}
		\tilde{t}&=
		\alpha^\prime \,\epsilon_{J/\psi i}\,\epsilon_{\gamma i}\,\epsilon_{\rho^0j}\,\epsilon_{\omega j}\\
		&+\beta^\prime \,\epsilon_{J/\psi i}\,\epsilon_{\rho^0i}\,\epsilon_{\gamma j}\,\epsilon_{\omega j}\\
		&+\gamma^\prime \,\epsilon_{J/\psi i}\,\epsilon_{\omega i}\,\epsilon_{\gamma j}\,\epsilon_{\rho^0j},
	\end{aligned}
\end{equation}
with
\begin{align}
		\alpha^\prime
		=&\frac{e}{g}\left\{
		\left(3A+2B\right)
		\left(
		1+\frac{5}{3}\,G_{\rho^0\omega}\,t_{\rho^0\omega,\rho^0\omega}(M_{\text{inv}}(\rho^0\omega))
		\right)  \right.\nonumber\\
		&+\sqrt{2}B\frac{5}{3}\,G_{\rho^0\phi}\,t_{\rho^0\phi,\rho^0\omega}(M_{\text{inv}}(\rho^0\omega))  \nonumber\\
		&+A\frac{5}{3}\,G_{K^{*+}K^{*-}}\,t_{K^{*+}K^{*-},\rho^0\omega}(M_{\text{inv}}(\rho^0\omega))  \nonumber\\
		&\left.-2A\frac{5}{3}\,G_{K^{*0}\bar{K}^{*0}}\,t_{K^{*0}\bar{K}^{*0},\rho^0\omega}(M_{\text{inv}}(\rho^0\omega))\right\},\label{eq:alpha_prime}
\end{align}
\begin{equation}
	\begin{aligned}
		\beta^\prime=\gamma^\prime=\frac{e}{g}\left(3A+2B\right),
	\end{aligned}\label{eq:beta_gamma_prime}
\end{equation}
where the corresponding transition amplitudes are given by
\begin{equation}
	t_{\rho^0\omega,\rho^0\omega}=\dfrac{g_{\rho\omega}g_{\rho\omega}}{M_{\text{inv}}^2(\rho^0\omega)-M_R^2+iM_R\Gamma_R},
\end{equation}
\begin{equation}
	t_{\rho^0\phi,\rho^0\omega}=\dfrac{g_{\rho\phi}g_{\rho\omega}}{M_{\text{inv}}^2(\rho^0\omega)-M_R^2+iM_R\Gamma_R},
\end{equation}
\begin{equation}
	t_{K^{*+}K^{*-},\rho^0\omega}=-\frac{1}{\sqrt{2}}\dfrac{g_{K^{*}\bar{K}^{*}}g_{\rho\omega}}{M_{\text{inv}}^2(\rho^0\omega)-M_R^2+iM_R\Gamma_R},
\end{equation}
\begin{equation}
	t_{K^{*0}\bar{K}^{*0},\rho^0\omega}=\frac{1}{\sqrt{2}}\dfrac{g_{K^{*}\bar{K}^{*}}g_{\rho\omega}}{M_{\text{inv}}^2(\rho^0\omega)-M_R^2+iM_R\Gamma_R}.
\end{equation}

Summing and averaging $|\tilde{t}|^2$ over the particle spins in the Coulomb gauge, $\epsilon^0 = 0$, $\vec{\epsilon} \cdot \vec{P}_\gamma = 0$, we obtain,
\begin{align}\label{eq:ttotal_raidative}
	&\overline{\sum}\sum|\tilde{t}|^2
	=\frac{1}{3}\left\{6\left|\alpha^\prime\right|^2+6\left|\beta^\prime\right|^2+6\left|\gamma^\prime\right|^2\right. \nonumber\\
	&\left.+2\times2\textbf{Re}(\alpha^\prime\beta^{\prime*})+2\times2\textbf{Re}(\beta^\prime\gamma^{\prime*})+2\times2\textbf{Re}(\alpha^\prime\gamma^{\prime*})\right\} \nonumber\\
	&=\left\{2\left|\alpha^\prime\right|^2+2\left|\beta^\prime\right|^2+2\left|\gamma^\prime\right|^2+\frac{4}{3}\textbf{Re}(\alpha^\prime\beta^{\prime*})+\frac{4}{3}\textbf{Re}(\beta^\prime\gamma^{\prime*})\right. \nonumber\\
	&\left.+\frac{4}{3}\textbf{Re}(\alpha^\prime\gamma^{\prime*})\right\}.
\end{align}
The factors 6 and 2 in Eq.~(\ref{eq:ttotal_raidative}), rather than 9 and 3 in Eq.~(\ref{eq:ttotal_strong}), arise because the sum is performed over transverse photon polarizations, with the property $\sum \epsilon_{\gamma i} \epsilon_{\gamma j} = (\delta_{ij} - \frac{P_{\gamma i} P_{\gamma j}}{\vec{P}_\gamma^2})$. 

\subsection{Determination of $A$ and $B$}
At this stage, the parameters $A$ and $B$ remain undetermined. However, experimental data on several radiative branching ratios can be used to constrain them. Using the same formalism discussed above, the squared amplitudes for the reactions $J/\psi \to \gamma \rho \rho$, $J/\psi \to \gamma \omega \omega$, $J/\psi \to \gamma \phi \phi$, $J/\psi \to \gamma K^{*} \bar{K}^{*}$, $J/\psi \to \gamma \rho \phi$ and $J/\psi \to \gamma \rho \omega$ can be obtained. Our purpose here is not to perform a precise study of all these reactions, but rather to obtain a rough estimate of $A$ and $B$, so that we can estimate the absolute rates and assess whether they are within the measurable range at BESIII. Note that we aim to study the feasibility of observing the $a_0(1710)$ line shapes in the mass distributions of $J/\psi \to \rho^+ \rho^- \omega$ and $J/\psi \to \gamma \rho^0 \omega$ which, as we shall see, do not depend on the precise values of $A$ and $B$. With this perspective, it is sufficient to study these reactions at the tree level for the determination of $A$ and $B$, but we implement the final state interaction in $J/\psi\to\gamma\rho\omega$ when evaluating the mass distributions to see the $a_0(1710)$. Using Eqs.~(\ref{eq:VVV}), (\ref{eq:VV_V}) and (\ref{eq:V_V_V}) again, along with the spin sums as previously evaluated, we find
\begin{equation}
	\begin{aligned}
		\overline{\sum}\sum|t_{\gamma \rho^+ \rho^-}|^2&=10\left(\frac{e}{g}\right)^2A^2,
	\end{aligned}
\end{equation}
\begin{equation}
	\begin{aligned}
		\overline{\sum}\sum|t_{\gamma \rho^0 \rho^0}|^2&=10\left(\frac{e}{g}\right)^2\frac{A^2}{2},
	\end{aligned}
\end{equation}
\begin{equation}
	\begin{aligned}
		\overline{\sum}\sum|t_{\gamma \omega \omega}|^2&=10\left(\frac{e}{g}\right)^2\frac{\left(A+\frac{4}{3}B\right)^2}{2},
	\end{aligned}
\end{equation}
\begin{equation}
	\begin{aligned}
		\overline{\sum}\sum|t_{\gamma \phi \phi}|^2&=10\left(\frac{e}{g}\right)^2\frac{\left(-2A-\frac{4}{3}B\right)^2}{2},
	\end{aligned}
\end{equation}
\begin{equation}
	\begin{aligned}
		\overline{\sum}\sum|t_{\gamma K^{*+} K^{*-}}|^2&=10\left(\frac{e}{g}\right)^2A^2,
	\end{aligned}
\end{equation}
\begin{table*}
\centering
\caption{Results for the radiative decays of several channels. In Fit 1 the rates of $\gamma \rho \phi$ and $\gamma \rho \omega$ are not considered. In the other fits different values of these rates are considered compatible with the experimental boundaries.}
\begin{tabular}{|c|c|c|c|c|c|}
	\hline
	Channel $J/\psi \to$ & Experiment $(\times 10^{-4})$ & \begin{tabular}{@{}c@{}}Fit 1 $(\times 10^{-4})$ \\ 5) - \\ 6) -\end{tabular} & \begin{tabular}{@{}c@{}}Fit 2 $(\times 10^{-4})$ \\ 5) $0.5 \pm 0.5$ \\ 6) $3 \pm 3$\end{tabular} & \begin{tabular}{@{}c@{}}Fit 3 $(\times 10^{-4})$ \\ 5) $0.7 \pm 0.3$ \\ 6) $4 \pm 2$\end{tabular} & \begin{tabular}{@{}c@{}}Fit 4 $(\times 10^{-4})$ \\ 5) $0.25 \pm 0.75$ \\ 6) $2 \pm 4$\end{tabular} \\
	\hline
	1)~$\gamma \rho \rho$ & $4.5 \pm 0.8$ & $22$ & $3.19$ & $2.73$ & $4.00$ \\
    2)~$\gamma \omega \omega$ & $16.1 \pm 3.3$ & $19$ & $0.09$ & $0.05$ & $0.13$ \\
	3)~$\gamma \phi \phi$ & $4.0 \pm 1.2$ & $1.37$ & $0.79$ & $0.63$ & $1.03$ \\
	4)~$\gamma K^* \bar{K}^*$ & $40.0 \pm 13$ & $52.9$ & $7.88$ & $6.50$ & $9.51$ \\
	5)~$\gamma \rho \phi$ & $< 0.88$ & $81.2$ & $0.83$ & $0.82$ & $0.97$ \\
	6)~$\gamma \rho \omega$ & $< 5.4$ & $13.7$ & $7.93$ & $6.29$ & $10.3$ \\
	\hline
	$A$ & & $-0.20$ & $-0.076$ & $-0.070$ & $-0.085$ \\
	$B$ & & $0.40$ & $0.040$ & $0.040$ & $0.043$ \\
	\hline
\end{tabular}

\label{table:radiative_decay_fit}
\end{table*}
\begin{table*}
\centering
\caption{The branching ratio for the $J/\psi \to \rho^+ \rho^- \omega$ reaction.}
\begin{tabular}{|c|c|c|c|c|}
	\hline
	&~Fit 1 &~Fit 2 &~Fit 3 &~Fit 4 \\
	\hline
	~~~$\mathcal{B}r(J/\psi\to\rho^+\rho^-\omega)$~(Only tree level)~~~&~~$30.2\%$~~	&~~$17.5\%$~~ &~~$13.9\%$~~ &~~$22.7\%$~~ \\
	~~~$\mathcal{B}r(J/\psi\to\rho^+\rho^-\omega)$~(+FSI)~~~&~~$294\%$~~	&~~$45.7\%$~~ &~~$38.4\%$~~ &~~$57.9\%$~~ \\
	\hline
\end{tabular}
\label{table:strong_decay_fit}
\end{table*}
\begin{figure*}
\subfigure[]{
	\centering
	\includegraphics[scale=0.65]{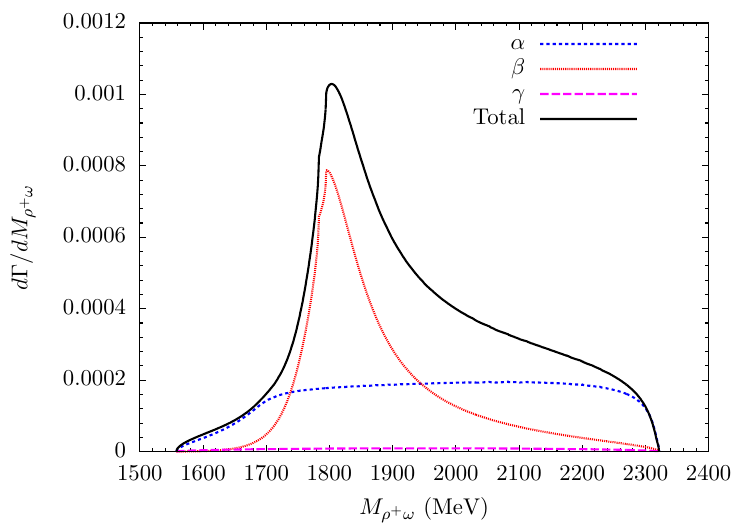}
	\label{fig:dwidth_rhop_omega_fit1}
}
\subfigure[]{
	\centering
	\includegraphics[scale=0.65]{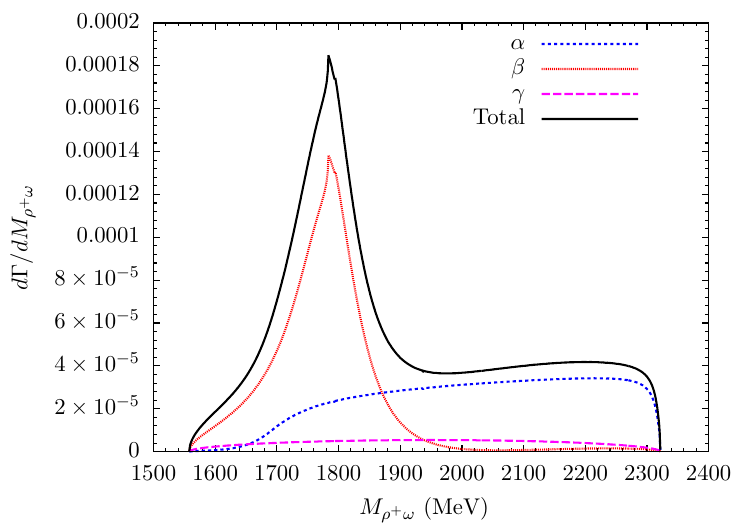}
	\label{fig:dwidth_rhop_omega_fit2}
}
\subfigure[]{
	\centering
	\includegraphics[scale=0.65]{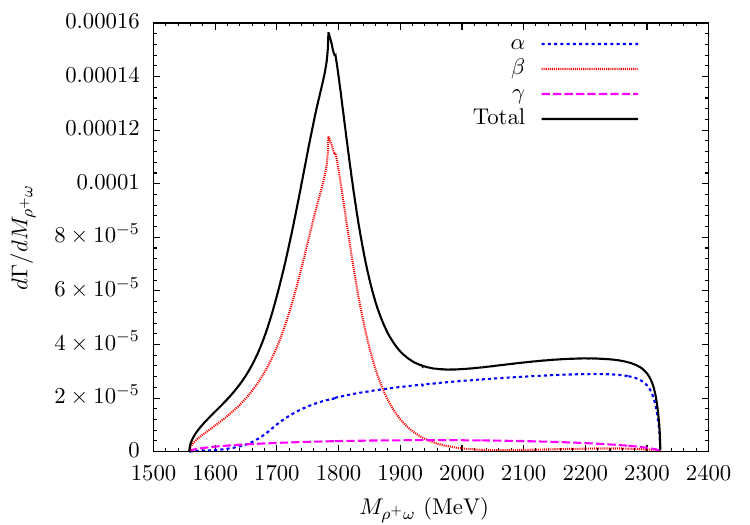}
	\label{fig:dwidth_rhop_omega_fit3}
}
\subfigure[]{
	\centering
	\includegraphics[scale=0.65]{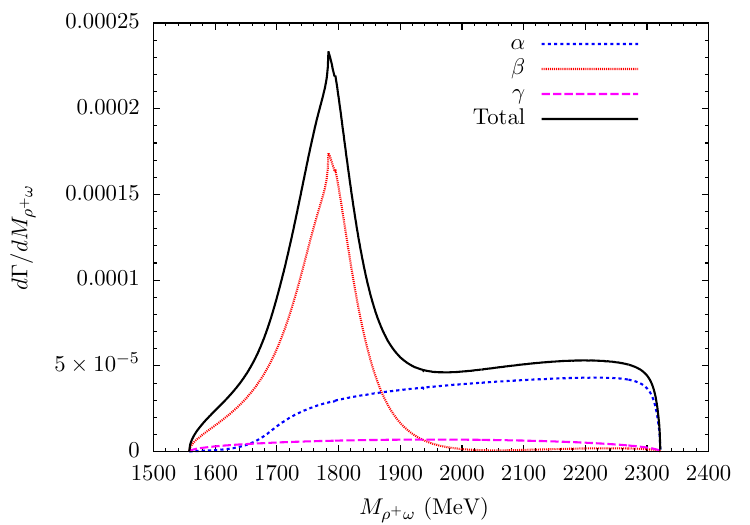}
	\label{fig:dwidth_rhop_omega_fit4}
}
\caption{$\rho^+ \omega$ mass distribution for $J/\psi \to \rho^+ \rho^- \omega$ for Fit 1 (a), Fit 2 (b), Fit 3 (c) and Fit 4 (d).}\label{fig:dwidth_rhop_omega}
\end{figure*}
\begin{equation}
	\begin{aligned}
		\overline{\sum}\sum|t_{\gamma K^{*0} \bar{K}^{*0}}|^2&=10\left(\frac{e}{g}\right)^2\left(-2A\right)^2,
	\end{aligned}
\end{equation}

\begin{equation}
	\begin{aligned}
		\overline{\sum}\sum|t_{\gamma \rho \phi}|^2&=10\left(\frac{e}{g}\right)^22B^2,
	\end{aligned}
\end{equation}
\begin{equation}
	\begin{aligned}
		\overline{\sum}\sum|t_{\gamma \rho \omega}|^2&=10\left(\frac{e}{g}\right)^2\left(3A+2B\right)^2.
	\end{aligned}
\end{equation}

By summing $K^{*+}K^{*-}$ and $K^{*0}\bar{K}^{*0}$, and $\rho^+\rho^-$ with $\rho^0\rho^0$, we find
\begin{equation}
	\begin{aligned}
		\overline{\sum}\sum|t_{\gamma K^{*} \bar{K}^{*}}|^2&=10\left(\frac{e}{g}\right)^25A^2,
	\end{aligned}
\end{equation}

\begin{equation}
	\begin{aligned}
		\overline{\sum}\sum|t_{\gamma \rho \rho}|^2&=10\left(\frac{e}{g}\right)^2\frac{3A^2}{2},
	\end{aligned}
\end{equation}

The factor $1/2$ for final production of two identical particles is already considered in the former equations.

\section{Results}\label{sec3}

\begin{figure*}[!t]
\subfigure[]{
		\centering
		\includegraphics[scale=0.65]{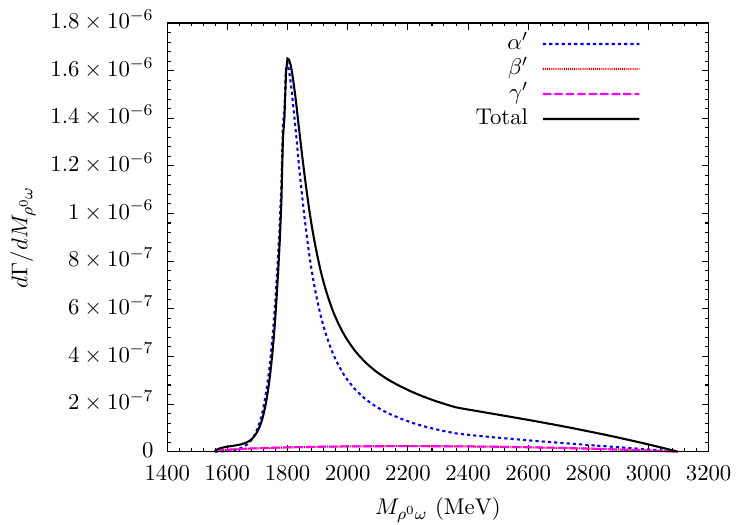}
		\label{fig:dwidth_rho0_omega_fit1}
}
\subfigure[]{
		\centering
		\includegraphics[scale=0.65]{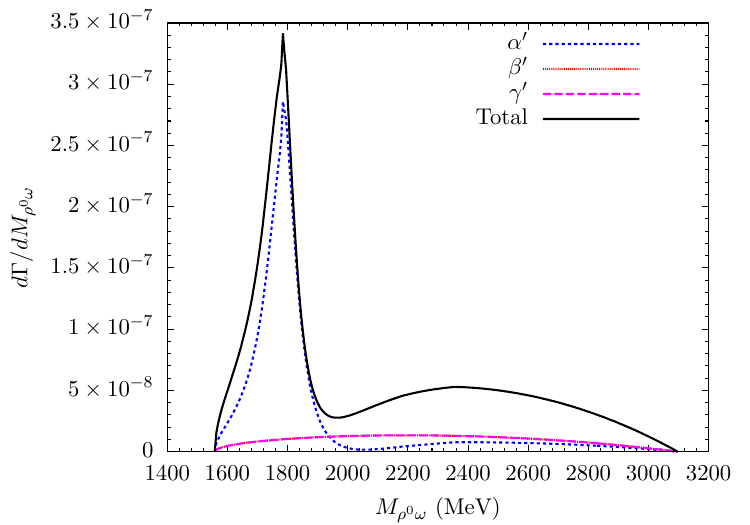}
		\label{fig:dwidth_rho0_omega_fit2}
}
\subfigure[]{
		\centering
		\includegraphics[scale=0.65]{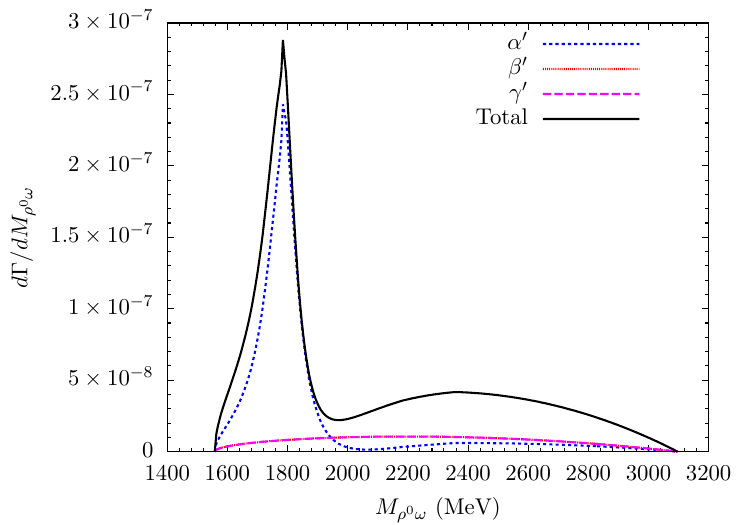}
		\label{fig:dwidth_rho0_omega_fit3}
}
\subfigure[]{
		\centering
		\includegraphics[scale=0.65]{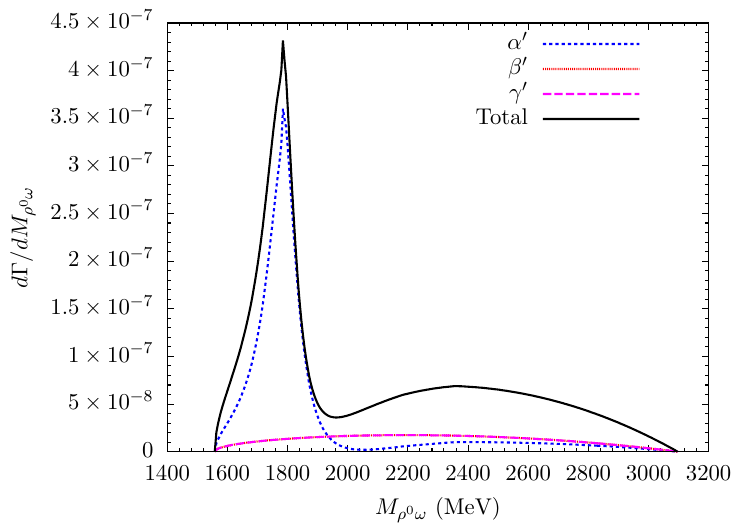}
		\label{fig:dwidth_rho0_omega_fit4}
}
    \caption{$\rho^0 \omega$ invariant mass distribution for Fit 1 (a), Fit 2 (b), Fit 3 (c) and Fit 4 (d).}
\end{figure*}

\begin{table}
\centering
\caption{Pole position and couplings with the simplified model of Eq.~\eqref{eq:BS_single_channel} with a single channel $K^*\bar{K}^*$ for different values of $q_{\text{max}}$.}
\begin{tabular}{c|c|c}
			\hline\hline
			~~~~$q_{\text{max}}$~[MeV]~~~~&~~~~Pole~[MeV]~~~~&~~~~$g$~[MeV]~~~~\\
			\hline
			900 & $1784$ & $4413.75$ \\
			\hline
			960 & $1780$ & $5754.96$ \\
			\hline
			1000 & $1776$ & $6564.42$ \\
			\hline\hline
		\end{tabular}
\label{table:new_pole_coupling}
\end{table}
\begin{figure*}
\subfigure[]{
	\centering
	\includegraphics[scale=0.65]{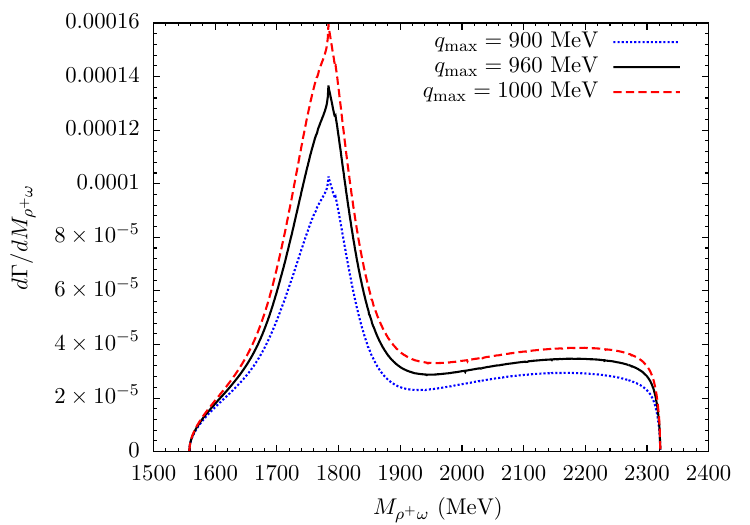}
}
\subfigure[]{
	\centering
	\includegraphics[scale=0.65]{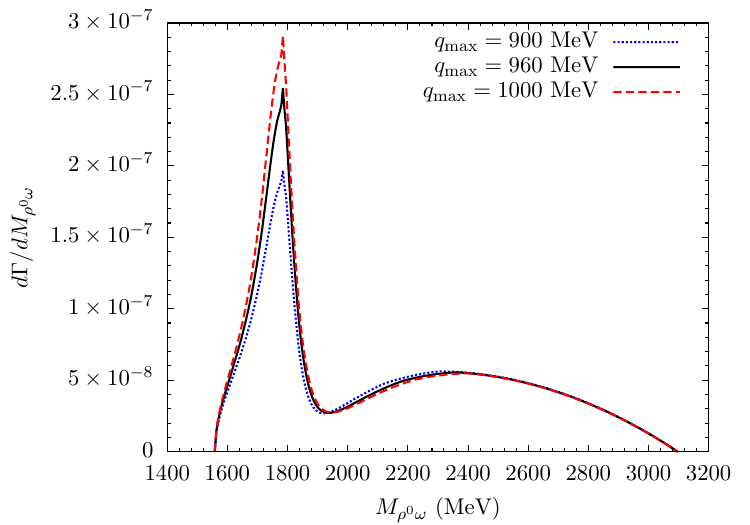}
}
\caption{(a) $\rho^+\omega$ mass distribution for $J/\psi\to\rho^+\rho^-\omega$ with different values of $q_{\text{max}}$, (b) the same for the $J/\psi\to\gamma\rho^0\omega$ reaction.}\label{fig:dwidth_uncertain}
\end{figure*}

We use the radiative decay data compiled by the Particle Data Book~\cite{ParticleDataGroup:2024cfk} for the reactions 1) $J/\psi \to \gamma \rho \rho$, 2) $J/\psi \to \gamma \omega \omega$, 3) $J/\psi \to \gamma \phi \phi$ and 4) $J/\psi \to \gamma K^{*} \bar{K}^{*}$ and fit the parameters $A$ and $B$. For the branching ratios of 5) $J/\psi \to \gamma \rho \phi$ and 6) $J/\psi \to \gamma \rho \omega$ there are only upper bounds available. We therefore perform several fits. In the first one (Fit 1) only the branching ratios for which central values are reported (the first four channels in Table~\ref{table:radiative_decay_fit}) are considered. In addition, we perform three more fits (Fits 2–4) in which we also include values below the experimental upper bounds for the $\gamma\rho\phi$ and $\gamma\rho\omega$ channels, exploring different choices compatible with these limits.  For each fit we evaluate the mass distributions for $J/\psi \to \gamma \rho^0 \omega$ and $J/\psi \to \rho^+ \rho^- \omega$ and in all of them we find a clear signal for the $a_0(1710)$ ($1780$~MeV in our approach), as we will show below. Although the uncertainties are sizable, our results show that the $a_0(1710)$ consistently appears as a prominent peak in the $\rho \omega$ mass distribution, and, also important, the branching ratios of the two reactions are large and comparable to measured branching fractions, two conditions that make the experimental study of these reactions advisable for obtaining a clear signal of the $a_0(1710)$ resonance.

The results of these fits are shown in Table~\ref{table:radiative_decay_fit}. The purpose of these fits is not to obtain a precise description of all radiative $J/\psi$ decay channels, but rather to determine reasonable ranges for the parameters $A$ and $B$ and estimate the order of magnitude of the corresponding branching fractions. The spread of the results obtained with the different fits provides an estimate of the model uncertainty associated with the determination of $A$ and $B$. In fact, as we will see below, this source of uncertainty is considerably larger than the statistical uncertainty that would be obtained within a given single fit and is expected to largely dominate over the uncertainties associated with the other parameters entering the calculation. For this reason, the results obtained with all four fits are retained throughout the analysis. A first observation from Table~\ref{table:radiative_decay_fit} is that Fit 1 does not provide a fully satisfactory description of the data. In particular, the $\gamma \rho \rho$ channel shows a significant discrepancy. This is not unexpected, since our calculation is performed at tree level, while the $\rho \rho$ interaction is known to be strong and generates resonances such as the $f_2(1270)$ and $f_0(1370)$~\cite{Molina:2008jw,Geng:2008gx}, which are not included here and can significantly modify the rate. However, we see that with the values of $A$ and $B$ obtained, the branching ratios for the $J/\psi \to \gamma \rho \phi$ and $J/\psi \to \gamma \rho \omega$ are bigger than the upper bounds. Another unsatisfactory feature is that $A$ is expected to be of the order of $3B$~\cite{Abreu:2023xvw,Molina:2019wjj} whereas in Fit 1 it is approximately $B/2$. As mentioned above, since final-state interactions between the two hadrons are not included in these estimates, the results should be regarded as indicative only. We then perform Fits 2-4, including values for the $\gamma \rho \phi$ and $\gamma \rho \omega$ channels compatible with the experimental upper limits. In these cases, the fitted values of $A$ and $B$ behave more naturally, with $A$ typically about twice $B$. However, this improvement comes at the cost of underestimating the measured branching ratios for several channels, in particular $ \gamma \omega \omega$, $\gamma\phi\phi$ and $\gamma K^* \bar{K}^*$. This is reflected in the smaller values of $A$ and $B$, which reduce all predicted rates. As mentioned above, one should not overinterpret the results obtained using only tree-level diagrams. However, the significant differences between the fits when imposing the constraints from the $J/\psi \to \gamma \rho \phi$ and $J/\psi \to \gamma \rho \omega$ channels suggest that improved measurements of these rates would be highly desirable.

While being open to future updates of these ratios, at present we keep in mind the large uncertainty of the results, but in all cases, the signal of the resonance is very clear, and the absolute rates are well within measurable capability of present facilities.

After this discussion, we come to our main purpose, which is to study the excitation of the $a_0(1710)$ resonance in the $J/\psi \to \gamma \rho^0 \omega$ and $J/\psi \to \rho^+ \rho^- \omega$ reactions. 
We first discuss the branching ratio for the $J/\psi \to \rho^+ \rho^- \omega$ reaction, shown in Table~\ref{table:strong_decay_fit}. For Fit 1, we obtain an unphysically large branching ratio, exceeding unity. However, as seen in Table~\ref{table:radiative_decay_fit}, this fit also overestimates the $J/\psi \to \gamma \rho \omega$ rate by roughly a factor of three. If we rescale the rate obtained by this factor, we obtain results roughly in line with those obtained with the other fits.

The general conclusion is that, despite the large uncertainties, the branching ratio for $J/\psi \to \rho^+ \rho^- \omega$ is consistently large, typically above 20\%, indicating that this channel should be prominent in $J/\psi$ decays. Similarly, the predicted branching ratios for $J/\psi \to \gamma \rho \omega$ are of the order of $(5\sim15)\times10^{-4}$, comparable to other measured radiative decay modes.

We now turn to the invariant mass distributions. Fig.~\ref{fig:dwidth_rhop_omega} shows the $\rho^+ \omega$ mass distribution for the $J/\psi \to \rho^+ \rho^- \omega$ reaction for the different fits. The four panels correspond to the four fits discussed above. Rather than displaying conventional statistical error bands, we show explicitly the results obtained with the different fits, since the spread among them provides a more realistic estimate of the uncertainty of the calculation. In this way, both the uncertainty in the overall strength and the possible variations in the shape of the invariant-mass distribution can be directly assessed. In all cases, a clear peak associated with the $a_0(1710)$ appears around 1800 MeV, standing out over a smooth background. We also show separately the contributions of the different terms discussed in the formalism. In Fig.~\ref{fig:dwidth_rhop_omega_fit1}, for Fit 1, the $\rho^+\omega$ mass distribution is decomposed into the $\alpha$, $\beta$, and $\gamma$ components of Eqs.~(\ref{eq:alpha}),~(\ref{eq:beta}) and~(\ref{eq:gamma}). The interpretation of these terms is straightforward. The $\gamma$ term corresponds purely to the tree-level contribution. The $\alpha$ term contains both the tree-level contribution and rescattering from vector--vector channels, but depends on the $\rho^-\omega$ invariant mass, which is not the one shown in the $\rho^+\omega$ distribution, and therefore does not generate any resonant structure in this case. In contrast, the $\beta$ term has a similar structure but depends on the $\rho^+\omega$ invariant mass. As a result, the resonance signal appears exclusively in this term, producing a pronounced peak that clearly dominates over the background from the other contributions.

Comparing the different fits, we observe that Fit 1 exhibits a substantially larger overall normalization than Fits 2-4 in Figs.~\ref{fig:dwidth_rhop_omega_fit2},~\ref{fig:dwidth_rhop_omega_fit3},~\ref{fig:dwidth_rhop_omega_fit4}, reflecting the larger values of the parameters $A$ and $B$ obtained in that fit. As discussed above, these values also lead to a poorer description of some radiative decay channels. Nevertheless, the resonant contribution associated with the $a_0(1710)$ shows a very similar shape in all fits. The main differences arise from the relative importance of the non-resonant background and the interference among the different terms of the amplitude. Consequently, while the overall normalization and background level vary, the position and visibility of the $a_0(1710)$ peak remain stable against variations of $A$ and $B$.

We now turn to the radiative $J/\psi\to\gamma\rho^0 \omega$ decay. The main features are similar to those found in the strong case. In Figs.~\ref{fig:dwidth_rho0_omega_fit1},~\ref{fig:dwidth_rho0_omega_fit2},~\ref{fig:dwidth_rho0_omega_fit3} and~\ref{fig:dwidth_rho0_omega_fit4}, corresponding to Fits 1-4, the $\rho^0 \omega$ invariant mass distribution exhibits a pronounced peak associated with the $a_0(1710)$, even more clearly separated from the background than in the strong decay. While the overall normalization changes from one fit to another, in accordance with Table~\ref{table:radiative_decay_fit}, the line shapes remain essentially unchanged, and the $a_0(1710)$ signal is clearly visible in all cases. It is worth noting that, according to Eqs.~(\ref{eq:alpha_prime}) and~(\ref{eq:beta_gamma_prime}), the resonance contribution originates exclusively from the $\alpha'$ term. Although the integrated branching ratio for this radiative decay carries sizable uncertainties, it is of the order of $10^{-4}$, making this reaction accessible to present experimental facilities and allowing for an improvement over the current upper bounds. 

To further estimate uncertainties of the results, we show their variations with moderate changes of the parameter $q_{\text{max}}$ which does not destroy the global fit to the different vector-vector molecular states studied in~\cite{Geng:2008gx}. For this, we choose now the values $q_{\text{max}} = 900$~MeV and $1000$~MeV in addition to the $q_{\text{max}} = 960$~MeV used in the former calculations. Since the purpose is to show that the uncertainties from this source are small compared with those of the different fits made above, we consider a simple model to determine these uncertainties. We take just the dominant component $K^*\bar{K}^*$ and then the $T$ matrix is given by
\begin{equation}\label{eq:BS_single_channel}
T = \frac{1}{V^{-1} - G}. 
\end{equation}
We take $G$ with $q_{\text{max}} = 960$~MeV and evaluate $V$ such as to get a pole at $1780$~MeV. We get $V = -98$~MeV. Then use this potential to evaluate $T$ of Eq.~\eqref{eq:BS_single_channel} for $q_{\text{max}} = 900$~MeV and $1000$ MeV. In Table~\ref{table:new_pole_coupling} we show the results for the new pole positions and the couplings obtained. We see that the changes in the mass are moderate, around $5$~MeV, and the couplings also change in about $25\%$. We should not worry about the discrepancy of the couplings obtained with this simplified approach with those of Table~\ref{table:coupling_geng}, which are a consequence of neglecting the coupled channels and box diagrams. The purpose here is only to show that the uncertainties here produce small variations in the evaluated mass distributions. For this, we show in Fig.~\ref{fig:dwidth_uncertain} the results for fit 2 and the different $q_{\text{max}}$ values. As we can see, the changes induced are of the order of $20\%$, both in the $J/\psi \to \rho^+ \rho^- \omega$ and $J/\psi \to \gamma \rho^0 \omega$ reactions. These changes are small compared with the uncertainties that we have from other sources discussed above.

What one must stress from the results and their uncertainties is that the peak of the resonance shows up clearly in all cases, which will make it possible to have a precise determination of the mass of the resonance, and the rates are sufficiently large to be measured in present facilities.

\section{ Conclusions }\label{sec4}
We investigate the strong decay $J/\psi\to\rho^+\rho^-\omega$ and the radiative decay $J/\psi\to\gamma\rho^0\omega$, focusing on the dynamical generation of the scalar resonance $a_0(1710)$ through $S$-wave vector-vector final state interactions. To constrain the production weights of the primary vertices, we performed multiple fits using existing experimental data for various $J/\psi$ radiative decays.

Our results show that both reactions exhibit a clear and pronounced peak in the $\rho \omega$ invariant mass distribution around 1.8 GeV, associated with the $a_0(1710)$ resonance, on top of a background produced by the tree level and non-resonant vector-vector interactions. This feature is remarkably stable against variations of the parameters within the range allowed by the fits, indicating that the shape of the distribution is largely independent of the precise values of these parameters.

We find that the branching ratio for the $J/\psi \to \rho^+ \rho^- \omega$ reaction is consistently large, typically above 20\%, making this channel a prominent one in $J/\psi$ decays. For the radiative decay $J/\psi \to \gamma \rho \omega$, the predicted branching ratios are of the order of $(5\sim15)\times10^{-4}$ and are within reach of current experimental facilities. The radiative decay provides a particularly clean environment, where the $a_0(1710)$ signal appears even more clearly separated from the background. This makes it especially suitable for experimental studies.

In view of these results, we strongly encourage future measurements of the $J/\psi \to \rho^+ \rho^- \omega$ and $J/\psi \to \gamma \rho^0 \omega$ reactions at BESIII~\cite{BESIII:2009fln}, Belle~II~\cite{BelleII}, and the future Super Tau-Charm Facility (STCF)~\cite{Petrov:2026nxj}. Observing the predicted peak structures would not only confirm the presence of the $a_0(1710)$ in these decay channels, but also provide crucial insights into its nature and allow a more precise determination of its mass, which remains rather uncertain at present.

\section*{Acknowledgments}
This work was supported by the National Key R\&D Program of China (Grant No. 2024YFE0105200), the Natural Science Foundation of Henan (Grant No. 252300423951), and the Zhengzhou University Young Student Basic Research Projects for PhD students (Grant No. ZDBJ202522). Wen-Tao Lyu acknowledges the support of the China Scholarship Council. 
This work is also partly supported by the Spanish Ministerio de Economia y Competitividad~(MINECO) and European FEDER funds under Contracts No. FIS2017-84038-C2-1-PB, PID2020-112777GB-I00, and by Generalitat Valenciana under contract PROMETEO/2020/023. This project has received funding from the European Union Horizon 2020 research and innovation program under the program H2020-INFRAIA-2018-1, grant agreement No. 824093 of the STRONG-2020 project.
Research partially supported by grant PID2023-147458NB-C21 funded by MCIN/AEI/ 10.13039/501100011033 and by the European Union.

\end{document}